\documentclass[aip,apl,amssymb,amsmath,preprint,floatfix,superscriptaddress,showpacs]{revtex4-1}
\usepackage{graphicx}

\begin{document}

\title{Laser-induced THz magnetization precession for a tetragonal Heusler-like nearly compensated ferrimagnet}

\author{S. Mizukami}
\email{mizukami@wpi-aimr.tohoku.ac.jp}
\author{A. Sugihara}
\affiliation{
WPI Advanced Institute for Materials Research,
Tohoku University, Sendai 980-8577, Japan}
\author{S. Iihama}
\author{Y. Sasaki}
\affiliation{Department of Applied Physics, Tohoku University, Sendai 980-8579, Japan}

\author{K. Z. Suzuki}
\author{T. Miyazaki}
\affiliation{
WPI Advanced Institute for Materials Research,
Tohoku University, Sendai 980-8577, Japan}

\date{\today}

\begin{abstract}
Laser-induced magnetization precessional dynamics was investigated in epitaxial films of Mn$_3$Ge, 
which is a tetragonal Heusler-like nearly compensated ferrimagnet.
The ferromagnetic resonance (FMR) mode was observed,
the precession frequency for which exceeded 0.5 THz and originated from the large magnetic anisotropy field of approximately 200 kOe for this ferrimagnet.
The effective damping constant was approximately 0.03.
The corresponding effective Landau-Lifshitz constant of approximately 60 Mrad/s
and is comparable to those of the similar Mn-Ga materials.
The physical mechanisms for the Gilbert damping and for the laser-induced excitation of the FMR mode were
 also discussed in terms of the spin-orbit-induced damping and the laser-induced ultrafast modulation of the magnetic anisotropy, respectively.
\end{abstract}


\maketitle
Among the various types of magnetization dynamics,
coherent magnetization precession, {\it i.e.,} ferromagnetic resonance (FMR), is the most fundamental type,
and plays a major role in rf spintronics applications based 
on spin pumping\cite{Silsbee1979,Mizukami2001,Urban2001,Mizukami2002,Tserkovnyak2002}
and the spin-transfer-torque (STT).\cite{Berger1996,Slonczewski1996}
Spin pumping is a phenomenon through which magnetization precession generates dc and rf spin currents 
in conductors that are in contact with magnetic films.
The spin current can be converted into an electric voltage through the inverse spin-Hall effect.\cite{Saitoh2006}
The magnitude of the spin current generated via spin pumping is proportional to the FMR frequency $f_{\rm FMR}$;\cite{Mizukami2002,Tserkovnyak2002}
thus, the output electric voltage is enhanced with increased $f_{\rm FMR}$.
In the case of STT oscillators and diodes,
the $f_{\rm FMR}$ value for the free layer of a given magnetoresistive devices primarily 
determines the frequency range for those devices.\cite{Kiselev2003,Tulapurkar2005}
An STT oscillator and diode detector at a frequency of approximately 40 GHz have already been demonstrated;\cite{Bonetti2009,Maehara2014,Naganuma2015}
therefore, one of the issues for consideration as regards practical applications is the possibility of increasing $f_{\rm FMR}$ to hundreds of GHz or to the THz wave range (0.1-3 THz).\cite{Hoefer2005,Bonetti2009}

One simple method through which $f_{\rm FMR}$ can be increased utilizes magnetic materials with large perpendicular magnetic anisotropy fields $H_k^{\rm eff}$ and small Gilbert damping constants $\alpha$.\cite{Rippard2010,Taniguchi2013,Naganuma2015}
This is because $f_{\rm FMR}$ is proportional to $H_k^{\rm eff}$
and, also, because the FMR quality factor and critical current of an STT-oscillator are inversely and directly proportional to $\alpha$, respectively.
The $H_k^{\rm eff}$ value is determined by the relation $H_k^{\rm eff}=2K_u/ M_s-4\pi M_s$ for thin films,
where $K_u$ and $M_s$ are the perpendicular magnetic anisotropy constant and saturation magnetization, respectively.
Thus, materials with a small $M_s$, large $K_u$, and low $\alpha$ are very favorable;
these characteristics are similar to those of materials used in the free layers of magnetic tunnel junctions integrated in gigabit STT memory application.\cite{Yamada2015}
We have previously reported that the Mn-Ga metallic compound satisfies the above requirements,
and that magnetization precession at $f_{\rm FMR}$ of up to 0.28 THz was observed in this case.\cite{Mizukami2011}
A couple of research groups have studied magnetization precession dynamics in the THz wave range for the FePt films with a large $H_k^{\rm eff}$,
and reported an $\alpha$ value that is a factor of about 10 larger than that of Mn-Ga.\cite{He13,Iihama13,Becker14}
Thus, it is important to examine whether there are materials exhibiting properties similar to those of Mn-Ga exist,
in order to better understand the physics behind this behavior. 

In this letter, we report on observed magnetization precession at $f_{\rm FMR}$ of more than 0.5 THz for an epitaxial film of a Mn$_3$Ge metallic compound.
Also, we discuss the relatively small observed Gilbert damping.
Such THz-wave-range dynamics can be investigated by means of a THz wave\cite{Nakajima2010} or pulse laser.
Here, we use the all-optical technique proposed previously;\cite{Kampen2002} 
therefore, the mechanism of laser-induced magnetization precession is also discussed,
because this is not very clearly understood.

Mn$_3$Ge has a tetragonal D0$_{22}$ structure,
and the lattice constants are $a=3.816$ and $c=7.261$ \AA\ in bulk materials  [Fig. 1(a)].\cite{Ohoyama1961,Kren1971}
The Mn atoms occupy at two non-equivalent sites in the unit-cell.
The magnetic moment of Mn$_\mathrm{I}$ ($\sim$ 3.0 $\mu_B$) is anti-parallel to that of Mn$_\mathrm{II}$ ($\sim 1.9$ $\mu_B$), because of anti-ferromagnetic exchange coupling, and the net magnetic moment is $\sim 0.8$ $\mu_B$/f.u.
In other words, this material is a nearly compensated ferrimagnet with a Curie temperature $T_\mathrm{c}$ over 800 K.\cite{Yamada1990}
The tetragonal structure induces a uniaxial magnetic anisotropy, where the $c$-axis is the easy axis.\cite{Kren1971}
The D0$_{22}$ structure is identical to that of tetragonally-distorted D0$_3$, 
which is a class similar to the L2$_1$ Heusler structure; thus, D0$_{22}$ Mn$_3$Ge is also known as a tetragonal Heusler-like compound, as is Mn$_3$Ga.\cite{Balke2007}
The growth of epitaxial films of D0$_{22}$ Mn$_3$Ge has been reported quite recently, 
with these films exhibiting a large $K_u$ and small $M_s$,
similar to Mn-Ga.\cite{Kurt2012,Mizukami2013,Sugihara2014}
Note that  Mn$_3$Ge films with a single D0$_{22}$ phase can be grown for near stoichiometric compositions. \cite{Mizukami2013,Sugihara2014}
Further, an extremely large tunnel magnetoresistance is expected in the magnetic tunnel junction with Mn$_3$Ge electrodes,
owing to the fully spin-polarized energy band with $\Delta_1$ symmetry and the Bloch wave vector parallel to the $c$-axis at the Fermi level.
\cite{Mizukami2013,Miura2014}
These properties constitute the qualitative differences between the Mn$_3$Ge and Mn$_3$Ga compounds from the material perspective.

All-optical measurement for the time-resolved magneto-optical Kerr effect 
was employed using a standard optical pump-probe setup with a Ti: sapphire 
laser and a regenerative amplifier.
The wavelength and duration of the laser pulse were approximately $800$ nm and $150$ fs, respectively,
while the pulse repetition rate was 1 kHz. 
The pulse laser beam was divided into an intense pump beam and a weaker probe beam; 
both beams were $s$-polarized.
The pump beam was almost perpendicularly incident to the film surface,
whereas the angle of incidence of the probe beam was $\sim 6^\circ$ with respect to the film normal [Fig. 1(b)].
Both laser beams were focused on the film surface and the beam spots were overlapped spatially.
The probe and pump beams had spot sizes with 0.6 and 1.3 mm, respectively.
The Kerr rotation angle of the probe beam reflected at the film surface was analyzed 
using a Wollaston prism and balanced photodiodes.
The pump beam intensity was modulated by a mechanical chopper at a frequency of 360 Hz.
Then, the voltage output from the photodiodes was detected using a lock-in amplifier,
as a function of delay time of the pump-probe laser pulses.
The pump pulse fluence was $\sim$0.6 mJ/cm$^2$.
Note that the weakest possible fluence was used in order to reduce the temperature increase
while maintaining the signal-to-noise ratio.
A magnetic field $H$ of 1.95 T with variable direction $\theta_H$ was applied using an electromagnet [Fig. 1 (b)].
\begin{figure}
\begin{center}
\includegraphics[width=6.5cm,keepaspectratio,clip]{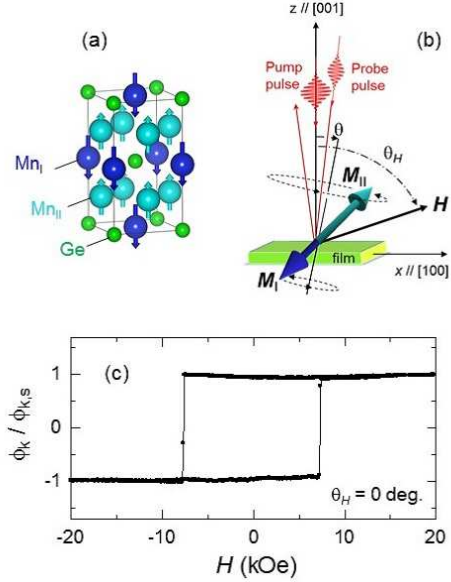}
\caption{
(a) Illustration of D0$_{22}$ crystal structure unit cell for Mn$_3$Ge.
(b) Diagram showing coordinate system used for optical measurement and ferromagnetic resonance mode of magnetization precession.
The net magnetization ($ = \mathbf{M}_{\rm II} - \mathbf{M}_{\rm I}$) precesses about the equilibrium angle of magnetization $\theta$,
where $\mathbf{M}_{\rm I}$ ($\mathbf{M}_{\rm II}$) is the magnetization vector for the Mn$_{\rm I}$ (Mn$_{\rm II}$)
sub-lattice.
(b) Out-of-plane normalized hysteresis loop of the Kerr rotation angle $\phi_k$ measured for the sample.
}
\end{center}
\end{figure}

The $c$-axis-oriented Mn$_3$Ge epitaxial films were grown on a single-crystalline (001) MgO substrate with a Cr seed layer, and were capped with thin MgO/Al layers at room temperature 
using a sputtering method with a base pressure below 1$\times10^{-7}$ Pa.
The characteristics of a 130-nm-thick film with slightly off-stoichiometric composition (74 at\% Mn) 
deposited at 500$^{\circ}$C are reported here,
because this sample showed the smallest coercivity (less than 1 T) and the largest saturation magnetization (117 
emu/cm$^3$) of a number of films grown with various thicknesses, compositions, and temperatures.
These properties are important to obtaining the data of time-resolved Kerr rotation angle $\phi_k$ 
with a higher signal-to-noise ratio,
because, as noted above, Mn$_3$Ge films have a large perpendicular magnetic anisotropy field and a small Kerr rotation angle.\cite{Sugihara2014}
Figure 1(c) displays an out-of-plane hysteresis loop of $\phi_k$ obtained for a sample
without pump-beam irradiation.
The loop is normalized by the saturation value $\phi_{k,s}$ at 1.95 T.
The light skin depth is considered to be about 30 nm for the employed laser wavelength, 
so that the $\phi_{k}$ value measured using the setup described above was almost proportional 
to the out-of-plane component of the magnetization $M_z$ within the light skin depth depth.
The loop shape is consistent with that measured using a vibrating sample magnetometer,
indicating that the film is magnetically homogeneous along the film thickness and that value of $\phi_{k}/\phi_{k,s}$ approximates to the $M_z$/$M_{\rm s}$ value.

Figure 2(a) shows the pump-pulse-induced change in the normalized Kerr rotation angle $\Delta \phi_{k}/\phi_{k,s}$ ($\Delta \phi_{k} = \phi_{k}-\phi_{k,s}$) as a function of the pump-probe delay time $\Delta t$ with an applied magnetic field $H$ perpendicular to the film plane.
$\Delta \phi_{k}/\phi_{k,s}$ decreases quickly immediately after the pump-laser pulse irradiation,
but it rapidly recovers within $\sim$2.0 ps.
This change is attributed to the ultrafast reduction and ps restoration of $M_s$ within the light skin depth region,
and is involved in the process of thermal equilibration among the internal degrees of freedom, {\it i.e.}, the electron, spin, and lattice systems.\cite{Beaurepaire1996}.
After the electron system absorbs light energy, the spin temperature increases in the sub-ps timescale because of the heat flow from the electron system, which corresponds to a reduction in $M_s$.
Subsequently, the electron and spin systems are cooled by the dissipation of heat into the lattices,
which have a high heat capacity.
Then, all of the systems reach thermal equilibrium.
This process is reflected in the ps restoration of $M_s$.
Even after thermal equilibrium among these systems is reached, 
the heat energy remains within the light skin depth region and the temperature is slightly higher than the initial value.
However, this region gradually cools via the diffusion of this heat deeper into the film
and substrate over a longer timescale.
Thus, the remaining heat causing the increased temperature corresponds to the small reduction of $\Delta \phi_{k}/\phi_{k,s}$ after $\sim$ 2.0 ps.

With increasing $\theta_H$ from out-of-plane to in-plane, 
a damped oscillation becomes visible in the $\Delta \phi_{k}/\phi_{k,s}$ data in the 2-12 ps range [Fig. 2(b)].
Additionally, a fast Fourier transform of this data clearly indicates a single spectrum at a frequency of 0.5-0.6 THz [Fig. 2(c)].
These damped oscillations are attributed to the temporal oscillation of $M_z$, 
which reflects the damped magnetization precession, \cite{Kampen2002}
because the $z$ component of the magnetization precession vector increases with increasing $\theta_H$.
Further, the single spectrum apparent in Fig. 2(c)
indicates that there are no excited standing spin-waves (such as those observed in thick Ni films),
even though the film is thicker than the optical skin depth.\cite{Kampen2002}
\begin{figure}
\begin{center}
\includegraphics[width=6.5cm,keepaspectratio,clip]{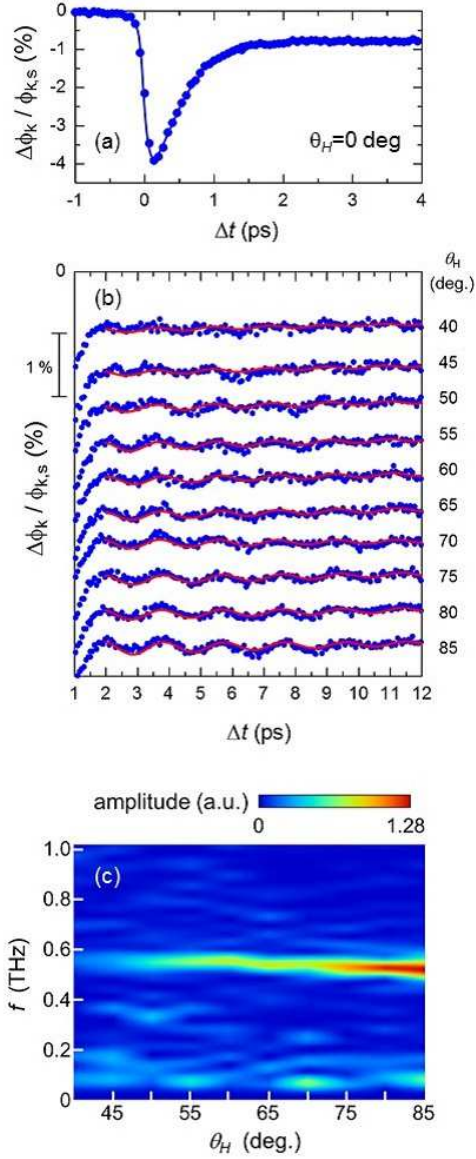}
\caption{
Change in Kerr rotation angle $\Delta \phi_k$ normalized by the saturation value $\phi_{k,s}$ 
as a function of pump-probe delay time $\Delta t$: (a) for a short time-frame at $\theta_H = 0^\circ$
and (b) for a relatively long time-frame and different values of $\theta_H$. 
The solid curves in (a) and (b) are a visual guide and values fitted to the data, respectively.
The data in (b) are plotted with offsets for clarity.
(c) Power spectral density as a function of frequency $f$ and magnetic field angle $\theta_H$.
}
\end{center}
\end{figure}

Ferrimagnets generally have two magnetization precession modes, {\it i.e.}, the FMR and exchange modes,
because of the presence of sub-lattices.\cite{Wangsness1953}
In the FMR mode, sub-lattice magnetization vectors precess
while maintaining an anti-parallel direction,
as illustrated in Fig. 1(b),
such that their frequency is independent of the exchange coupling energy between the sub-lattice magnetizations.
On the other hand, the sub-lattice magnetization vectors are canted in the exchange mode;
therefore, the precession frequency is proportional to the exchange coupling energy between them and
is much higher than that of the FMR mode.
As observed in the case of amorphous ferrimagnets,
the FMR mode is preferentially excited when the pump laser intensity is so weak that the increase in temperature is lower than the ferrimagnet compensation temperature.\cite{Mekonnen2011}
No compensation temperature is observed in the bulk Mn$_3$Ge. \cite{Ohoyama1961,Yamada1990}
Also, the temperature increase in this experiment is significantly smaller than $T_c$ 
because the reduction of $M_s$ is up to 4 \%, as can be seen in Fig. 2(a).
Therefore, the observed magnetization precession is attributed to the FMR mode.
Further, as the mode excitation is limited to the light skin depth,
the amplitude, frequency, and etc., for the excited mode are dependent on the film thickness with respect 
to the light skin depth.
This is because the locally excited magnetization precession propagates more deeply into the film as a spin wave in cases where $f_{\rm FMR}$ is in the GHz range.\cite{Kampen2002}
Note that it is reasonably assumed that such a non-local effect is negligible in this study,
because the timescale of the damped precession discussed here ($\sim$1-10 ps) is significantly shorter than that relevant to a spin wave with wavelength comparable to the light skin depth ($\sim$100 ps).

The FMR mode in the THz-wave range is quantitatively examined below.
When the exchange coupling between the sub-lattice magnetizations is sufficiently strong and the temperature is well below both $T_c$ and the compensation temperature,
the magnetization dynamics for a ferrimagnet can be described using the effective Landau-Lifshitz-Gilbert equation\cite{Mansuripur1995}
\begin{equation}
\frac{d\mathbf{m}}{dt} = - \gamma_\textmd{eff} \mathbf{m} \times \left[ \mathbf{H} + H_k^\textmd{eff}(\mathbf{m} \cdot \mathbf{z}) \mathbf{z} \right] + \alpha_\textmd{eff} \mathbf{m} \times \frac{d\mathbf{m}}{dt},
\end{equation}
where $\mathbf{m}$ is the unit vector of the net magnetization parallel (anti-parallel) to the magnetization vector $\mathbf{M}_{\rm II}$ ($\mathbf{M}_{\rm I}$) for the Mn$_{\rm II}$ (Mn$_{\rm I}$) sub-lattice [Fig. 1(b)].
Here, the spatial change of $\mathbf{m}$ is negligible, as mentioned above.
$H_k^{\rm eff}$ is the effective value of the perpendicular magnetic anisotropy field including the demagnetization field,
even though the demagnetization field is negligibly small for this ferrimagnet ($4\pi M_s = 1.5$ kOe).
Further, $\gamma_\textmd{eff}$ and $\alpha_\textmd{eff}$ are the effective values of the gyromagnetic ratio and the damping constant, respectively, 
which are defined as $\gamma_{\rm eff}=(M_{\rm II}-M_{\rm I})/(M_{\rm II}/\gamma_{\rm II}-M_{\rm I}/\gamma_{\rm I})$ and $\alpha_{\rm eff}=(\alpha_{\rm II}M_{\rm II}/\gamma_{\rm II}-\alpha_{\rm I}M_{\rm I}/\gamma_{\rm I})/(M_{\rm II}/\gamma_{\rm II}-M_{\rm I}/\gamma_{\rm I})$, respectively, using the gyromagnetic ratio $\gamma_{\rm I(II)}$ and damping constant $\alpha_{\rm I(II)}$ for the sub-lattice magnetization of Mn$_{\rm I(II)}$.
In the case of $H_k^{\rm eff} \gg H$, $f_{\rm FMR}$ and the relaxation time of the FMR mode $\tau_{\rm FMR}$ are derived from Eq. (1) as
\begin{eqnarray}
f_{\rm FMR} &=& \gamma_\textmd{eff} /2\pi \left( H_k^{\rm eff} + H_{z} \right), \\
1/\tau_{\rm FMR} &=& 2 \pi \alpha _{\rm eff} f_{\rm FMR}.
\end{eqnarray}
Here, $H_{z}$ is the normal component of $H$.
Figure 3(a) shows the $H_{z}$ dependence of the precession frequency $f_p$.
This is obtained using the experimental data on the oscillatory part of the change in $\Delta \phi_{k}/\phi_{k,s}$ via least-square fitting 
to the damped sinusoidal function, $\Delta \phi_{k,p}/\phi_{k,s} \exp(-t/\tau_p) \sin(2 \pi f_p + \phi_p)$, with an offset approximating the slow change of $\Delta \phi_{k}/\phi_{k,s}$ [solid curves, Fig. 2(b)].
Here, $\Delta \phi_{k,p}/\phi_{k,s}$, $\tau_p$, and $\phi_p$ are the normalized amplitude, relaxation time, and phase for the oscillatory part of $\Delta \phi_{k}/\phi_{k,s}$, respectively.
The least-square fitting of Eq. (2) to the $f_p$ vs. $H_z$ data yields $\gamma_{\rm eff}/2\pi$ = 2.83 GHz/kOe and $H_k^{\rm eff}$ = 183 kOe [solid line, Fig. 3(a)].
The $\gamma_{\rm eff}$ value is close to 2.80 GHz/kOe for the free electron.
The value of $H_k^{\rm eff}$ is equal to the value determined via static measurement (198 kOe) \cite{Sugihara2014} within the accepted range of experimental error.
Thus, the analysis confirms that the THz-wave range FMR mode primarily results from the large magnetic anisotropy field in the Mn$_3$Ge material.
The $\alpha_{\rm eff}$ values, which are estimated using the relation $\alpha_{\rm eff}=1/2\pi f_p \tau _p$ following Eq. (3),
are also plotted in Fig. 3(a).
The experimental $\alpha_{\rm eff}$ values are independent of $H_z$ within the accepted range of experimental error, 
being in accordance with Eq. (3); the mean value is 0.03.
This value of $\alpha_{\rm eff}$ for D0$_{22}$ Mn$_3$Ge is slightly larger than the previously reported values for
for D0$_{22}$ Mn$_{2.12}$Ga ($\sim$0.015) and L1$_0$ Mn$_{1.54}$Ga ($\sim$0.008).\cite{Mizukami2011}

In the case of metallic magnets, the Gilbert damping at ambient temperature is primarily caused by phonon and atomic-disorder scattering for electrons at the Fermi level in the Bloch states that are perturbed by the spin-orbit interaction. 
This mechanism, the so-called Kambersky mechanism,\cite{Kambersky1970,Kambersky1984} predicts $\alpha \propto M_s^{-1}$,
so that it is more preferable to use the Landau-Lifshitz constants $\lambda$ ($\equiv \alpha_{\rm} \gamma_{\rm} M_s$) for discussion of the experimental values of $\alpha$ for different materials.
Interestingly, $\lambda_{\rm eff}$ ($\equiv \alpha_{\rm eff} \gamma_{\rm eff} M_s$) for Mn$_3$Ge was estimated to be 61 Mrad/s, which is almost identical to the values for D0$_{22}$ Mn$_{2.12}$Ga ($\sim$ 81 Mrad/s) and L1$_0$ Mn$_{1.54}$Ga ($\sim$ 66 Mrad/s).
The $\lambda$ for the Kamberky mechanism is approximately proportional to $\lambda_{\rm SO}^2 D(E_F)$, 
where $\lambda_{\rm SO}$ is the spin-orbit interaction constant and $D(E_F)$ is the total density of states at the Fermi level.\cite{Kambersky1984}
The theoretical values of $D(E_F)$ for the above materials are roughly identical,
because of the similar crystal structures and constituent elements, 
even though the band structures around at the Fermi level differ slightly, as mentioned at the beginning.\cite{Mizukami2011,Mizukami2013}
Furthermore, the spin-orbit interactions for Ga or Ge, depending on the atomic number, may not differ significantly.
Thus, the difference in $\alpha_{\rm eff}$ for these materials can be understood qualitatively in terms of the Kambersky mechanism.
Further discussion based on additional experiments is required in order to obtain more precise values for $\alpha_{\rm eff}$ 
and to examine whether other relaxation mechanisms, such as extrinsic mechanisms (related to the magnetic inhomogeneities), 
must also be considered.
\begin{figure}
\begin{center}
\includegraphics[width=7cm,keepaspectratio,clip]{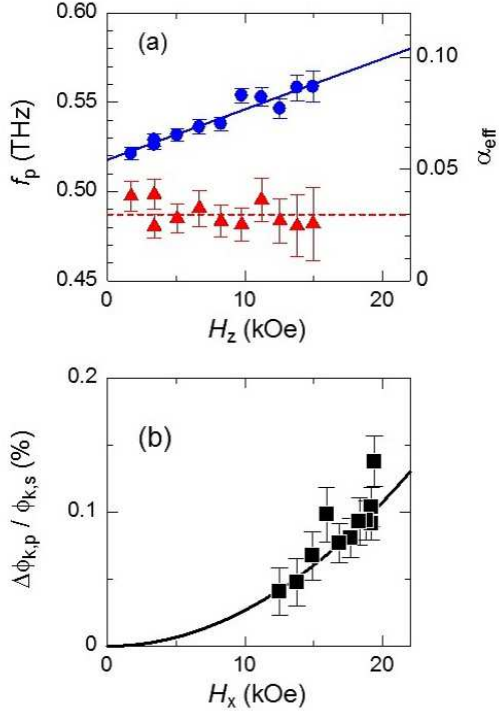}
\caption{
(a) Normal component of magnetic field $H$ dependence on precession frequency $f_p$ and effective damping constant $\alpha_{\rm eff}$ for Mn$_3$Ge film.
(b) Oscillation amplitude of the Kerr rotation angle $\Delta \phi_{k,p}/\phi_{k,s}$ corresponding to the magnetization precession
as a function of the in-plane component of $H$.
The solid line and curve are fit to the data. 
The dashed line denotes the mean value of $\alpha_{\rm eff}$.
}
\end{center}
\end{figure}

Finally, the excitation mechanism of magnetization precession in this study is discussed below,
in the context of a previously proposed scenario for laser-induced magnetization precession in Ni films. \cite{Kampen2002}
The initial equilibrium direction of magnetization $\theta$ is determined by the balance between $H$ and $H_k^{\rm eff}$ [Fig. 1(b)].
During the period in which the three internal systems are not in thermal equilibrium for $\Delta t < \sim$2.0 ps after the pump-laser irradiation [Fig. 2(a)],
not only the value of $M_s$, but also the value of the uniaxial magnetic anisotropy, {\it i.e.}, $H_k^{\rm eff}$, is altered.
Thus, the equilibrium direction deviates slightly from $\theta$ and is restored,
which causes magnetization precession.
This mechanism may be examined by considering the angular dependence of the magnetization precession amplitude.
Because the precession amplitude may be proportional to an impulsive torque generated from the modulation of $H_k^{\rm eff}$ in Eq. (1),
the torque has the angular dependence $| \mathbf{m}_0 \times \left( \mathbf{m}_0 \cdot \mathbf{z} \right) \mathbf{z} |$, 
where $\mathbf{m}_0$ is the initial direction of the magnetization.
Consequently, the $z$-component of the precession amplitude, {\it i.e.}, $\Delta \phi_{k,p}/\phi_{k,s}$, 
is expressed as
$\Delta \phi_{k,p}/\phi_{k,s} = \zeta \cos \theta \sin ^2 \theta \sim \zeta \left( H_{x}/H_k^{\rm eff} \right)^2$,
where $\zeta$ is the proportionality constant and $H_x$ is the in-plane component of $H$.
The experimental values of $\Delta \phi_{k,p}/\phi_{k,s}$ are plotted as a function of $H_{x}$ in Fig. 3(b).
The measured data match the above relation, 
which supports the above-described scenario.
Although $\zeta$ could be determined via the magnitude and the period of modulation of $H_k^{\rm eff}$,
it is necessary to consider the ultrafast dynamics of the electron, spin, and lattice in the non-equilibrium state
in order to obtain a more quantitative evaluation;\cite{Koopmans2010,Atxitia2012} 
this is beyond the scope of this report.

In summary, magnetization precessional dynamics was studied in a D0$_{22}$ Mn$_3$Ge epitaxial film using an all-optical pump-probe technique.
The FMR mode at $f_{\rm FMR}$ up to 0.56 THz was observed, which was caused by the extremely large  $H_k^{\rm eff}$.
A relatively small damping constant of approximately 0.03 was also obtained,
and the corresponding Landau-Lifshitz constant for Mn$_3$Ge were shown to be almost identical to that for Mn-Ga,
being in qualitatively accordance with the prediction of the Kambersky spin-orbit mechanism.
The field dependence of the amplitude of the laser-induced FMR mode was qualitatively consistent with the model based on the ultrafast modulation of magnetic anisotropy.

This work was supported in part by a Grant-in-Aid for Scientific Research 
for Young Researchers (No. 25600070) and for Innovative Area (Nano Spin Conversion Science, No. 26103004), 
NEDO, and the Asahi Glass Foundation.

\end{document}